\documentstyle[12pt]{article}
\begin{document}
\def\thefootnote{\fnsymbol{footnote}}
\begin{flushright}
KANAZAWA-07-08  \\ 
April, 2007
\end{flushright}
\vspace*{2cm}
\begin{center}
{\LARGE\bf  Phenomenological features in a model with non-universal 
gaugino CP phases}\\
\vspace{1 cm}
{\Large Daijiro Suematsu}
\footnote[1]{e-mail: suematsu@hep.s.kanazawa-u.ac.jp}
\vspace*{1cm}\\
{\it Institute for Theoretical Physics, Kanazawa University,\\
        Kanazawa 920-1192, Japan}\\    
\end{center}
\vspace{1cm}
{\Large\bf Abstract}\\  
We study phenomenological features in an extended gauge
mediation SUSY breaking model which has non-universal gaugino masses and
CP phases. We show that large CP phases in soft SUSY breaking 
parameters can be consistent with the constraints coming from 
the electric dipole moment (EDM) of an electron, a neutron, and also
a mercury atom. Masses of the superpartners are not necessarily required to be 
larger than 1~TeV but allowed to be $O(100)$~GeV. We also investigate 
the mass spectrum of Higgs scalars and their couplings to gauge 
bosons in that case. Compatibility of this model with the present 
experimental data on the Higgs sector is discussed.
\newpage
\setcounter{footnote}{0}
\def\thefootnote{\arabic{footnote}}
\section{Introduction}
In supersymmetric extensions of the standard model, new CP phases are
generally introduced through supersymmetry (SUSY) breaking.
Although these CP phases could play an interesting phenomenological role
related to the cosmological baryon number asymmetry, for example, 
it is well known that the electric dipole moment 
(EDM) of an electron and a neutron \cite{edmex} imposes 
severe constraints on such CP phases of soft SUSY 
breaking parameters in the minimal supersymmetric standard model
(MSSM) \cite{edmth,ko}. It seems to be very important to examine these
constraints because of their phenomenological consequences.  

Some possibilities to overcome these constraints have been proposed by 
now. In the first type solution, the soft SUSY breaking parameters 
are taken to be $O(100)$~GeV by assuming that the soft CP phases are 
smaller than $10^{-2}$ \cite{edmth}.
Since such small phases are not
protected by any symmetry, it is usually considered 
to be unnatural, and regarded as a CP problem in the MSSM.
In the second one, the soft CP phases are supposed to be $O(1)$ 
while a part of the relevant soft SUSY breaking parameters are assumed
to be $O(1)$~TeV or larger.\footnote{Various possibilities have been 
suggested. In one possibility, it is assumed
that the sfermions in the first and second generation have heavy masses of 
$O(1)$~TeV \cite{ko}. In another one, the $A$ parameters are assumed to be 
non-universal and those related to the first and second generation are 
supposed to be very small such as $A_f=(0,0,A)$ \cite{apara}. 
In this case, one needs to assume $\rm
arg(\mu)<10^{-2}$ and then the smallness of the CP phase is partially
required as in the first solution \cite{nonuni}.}
However,  considering the SUSY breaking larger than $O(1)$~TeV seems to be 
unattractive from a viewpoint of the weak scale SUSY. 
It may also be difficult
to expect any phenomenological effects through 
the present and near future experiments in this case.

As the third possibility, we can expect the cancellation among various
contributions to the EDMs \cite{cancel1,cancel2,gluino,gphase}.
If such a cancellation occurs and both the CP phases of $O(1)$
and the soft SUSY breaking parameters of $O(100)$~GeV can be 
consistent with the EDM constraints, we might have a lot of interesting 
phenomenology at the weak scale 
\cite{gphase,flav,cphiggs0,cphiggs1,cphiggs,Higgs}.
If we consider the origin of the baryon number asymmetry in the
universe due to electroweak baryogenesis, for example, it will be necessary 
to introduce some new sources of CP violation.
It is known that the Cabibbo-Kobayashi-Maskawa (CKM) phase in the 
standard model (SM) is insufficient to explain the baryon number asymmetry 
because of a suppression due to the smallness of the quark flavor mixing
\cite{ckm}.
If there exist large CP phases in the soft SUSY breaking parameters, 
the requirement for the electroweak phase transition to be 
strongly first order might be relaxed 
and the required Higgs mass bound could be larger \cite{ewbg}.
Various SUSY leptogenesis scenarios also seem to require the large CP 
phases in the soft SUSY breaking parameters \cite{nontherm,soft}.
Thus, the existence of such CP phases is a fascinating possibility 
from a viewpoint that they present us promising sources for 
the CP violation required in baryogenesis and leptogenesis.
Moreover, such CP phases might be checked through the LHC experiments.

Various works on this third possibility have suggested 
that the constraints on the EDMs of an
electron and a neutron could be satisfied even in the case that 
the CP phases in the soft SUSY breakings are $O(1)$ and the superpartners
are rather light.
It is based on the effective cancellation among various contributions 
to the EDMs.\footnote{In the case of the EDM of the electron, 
the cancellation between the chargino contribution and 
the neutralino contribution has been shown to occur \cite{cancel1,
cancel2,gphase}. On the other hand, the EDM of the neutron (EDMN) 
it has been known that there are several types of cancellation , 
that is, the cancellation between the diagrams of the gluino exchange 
and the chargino exchange 
diagrams and also the cancellation among the gluino exchange diagrams 
themselves {\it etc} \cite{cancel1,gluino}. In the case of the EDMN,
the combined effect of these cancellations allows 
the large soft CP phases \cite{cancel1,cancel2,gphase}. }
On the other hand, there is another claim that if we add the 
constraint from the EDM of the mercury atom, the allowed parameter 
regions disappear. It suggests that the parameter region for their 
cancellation are different between the electron and the mercury \cite{hg}.
However, it is useful to note that the usual analyses of the EDMs are 
based on the assumption for the universal gaugino masses as stressed 
in \cite{gphase}. If we do not take this assumption, we may find the 
way out of this difficulty.

Since the gaugino masses are universal in the usual SUSY breaking scenario,
it may be considered that such an assumption is unrealistic.
However, non-universal gaugino masses can be realized naturally, 
if we consider, for example, the intersecting D-brane 
model \cite{gphase,univ}, the extended gauge mediation SUSY 
breaking \cite{sue}, and the SUSY breaking mediated by the 
Abelian gaugino kinetic term mixing \cite{kinet}. 
In the previous paper \cite{st}, we examined the possibility of the
reconciliation between the CP phases of $O(1)$
and the experimental EDM constraints in a model with 
non-universal gaugino masses.
In that study we showed that the EDM constraints could be 
satisfied in rather large regions of the SUSY breaking 
parameter space under the existence of the large CP phases, 
as long as there are physical CP phases in the gaugino masses.
However, since the allowed parameter regions tend to be obtained for the 
small $\tan\beta$ \cite{st},  Higgs phenomenology might 
constrain the model strongly through the present 
Higgs search \cite{cphiggs0,cphiggs1,cphiggs,hgsearch}.

In this paper we extend the study to the Higgs sector using 
the parameter regions allowed by the 
constraints from the EDMs of the electron, the neutron, and the mercury
atom. We discuss the consistency of the scenario with 
the Higgs phenomenology.
The paper is organized as follows. In section 2  
we introduce the model for the soft SUSY breaking with 
the non-universal gaugino masses.
In section 3 we briefly describe the EDM of the mercury atom as an example
of the EDM calculation. The numerical analysis of the EDM constraints
is carried out by using the renormalization group study.
We apply this result to the estimation of the masses of the neutral
Higgs scalar and the couplings between the Higgs scalars and the gauge bosons. 
We also discuss predicted values of $g-2$ of the muon and the electron.
Section 4 is devoted to the summary.

\section{A model with non-universal gaugino CP phases}
We briefly introduce the model with non-universal gaugino masses 
studied in this paper and fix the notation.
We consider an extension of the well known minimal gauge mediation SUSY
breaking (GMSB) scenario,
which is defined by the following superpotential for the messenger
fields \cite{sue}:
\begin{equation}
W_m=\lambda_q\hat S_1\hat{\bar q}\hat q
+\lambda_\ell \hat S_2 \hat{\bar\ell}\hat\ell,
\end{equation}
where $\hat q,~\hat{\bar q}$ are ${\bf 3},~{\bf 3}^\ast$ of SU(3)$_c$ and
$\hat{\ell},~\hat{\bar\ell}$ are the doublets of SU(2)$_L$.
If both the singlet fields $\hat S_1$ and $\hat S_2$ couple with the 
hidden sector where the SUSY breaks down,
$\hat q, \hat{\bar q}$ and $\hat\ell, \hat{\bar\ell}$ 
play the role of messenger fields as
in the case of the ordinary scenario \cite{mgm1,mgm2,gmsbrev}.
Only difference from the ordinary minimal GMSB scenario is that 
$\hat q,~\hat{\bar q}$ and $\hat\ell,~\hat{\bar\ell}$ 
couple with the different singlet chiral 
superfields $\hat S_1$ and $\hat S_2$ in the superpotential $W_m$.
It is realized if we impose a suitable discrete symmetry on the 
model \cite{sue}.  If both their scalar components $S_\alpha$ and their
auxiliary components $F_{S_\alpha}$ 
obtain vacuum expectation values (VEVs) due to the couplings 
with the SUSY breaking sector, the masses of the gauginos and 
the scalars in the MSSM are 
generated at one-loop and two-loop level, respectively.  
They are represented as functions of 
$\Lambda_\alpha=\langle F_{S_\alpha}\rangle/\langle S_\alpha\rangle $
in a similar way as the ordinary scenario .
However, the mass formulas are somewhat modified from the usual ones
since the messenger fields $(\hat q,~\hat{\bar q})$ 
and $(\hat\ell,~\hat{\bar\ell})$ 
couple with the different singlets.

In this kind of model, the gaugino masses can be written in the form as
\cite{sue} 
\begin{equation}
M_3={\alpha_3\over 4\pi}\Lambda_1, \qquad
M_2={\alpha_2\over 4\pi}\Lambda_2, \qquad
M_1={\alpha_1\over 4\pi}\left({2\over 3}\Lambda_1+\Lambda_2\right), 
\label{eqff}
\end{equation}
where $\alpha_r=g^2_r/4\pi$ and $g_r$ stands for the coupling constant
for the standard model gauge group.
These formulas show that $M_3$ can be smaller 
than $M_{1,2}$ in the case of $\Lambda_2>\Lambda_1$. 
Since $\Lambda_\alpha$ is generally independent, the phases contained in the 
gaugino masses are non-universal even in the case of 
$\vert\Lambda_1\vert=\vert\Lambda_2\vert$. 
In that case, we cannot remove them completely 
by using the $R$-transformation unlike in the case of universal 
gaugino masses. In fact, if we define the phases as
$\Lambda_\alpha\equiv\vert\Lambda_\alpha\vert e^{i\theta_\alpha}$ and make
$M_2$ real by the $R$-transformation, the phases of the gaugino masses 
$M_r$ can be written as \cite{sue}
\begin{eqnarray}
&&\phi_3\equiv{\rm arg}(M_3)=\theta_1-\theta_2, \qquad 
\phi_2\equiv{\rm arg}(M_2)=0, \nonumber \\ 
&&\phi_1\equiv{\rm arg}(M_1)=\arctan\left({2\vert\Lambda_1\vert
\sin(\theta_1-\theta_2)\over
3\vert\Lambda_2\vert+2\vert\Lambda_1\vert\cos(\theta_1-\theta_2)}\right).
\label{eqfff}
\end{eqnarray}
These formulas show that the phases of the gaugino masses can be
parameterized three parameters, that is, $|\Lambda_1|$, $|\Lambda_2|$
and $\theta_1-\theta_2$. 

The scalar masses are induced through the two-loop diagrams as in 
the ordinary case.
Their formulas can be given as \cite{sue}
\begin{equation}
\tilde m^2_f=2\vert\Lambda_1\vert^2
\left[C_3\left({\alpha_3\over 4\pi}\right)^2 
+{2\over 3}\left({Y\over 2}\right)^2\left({\alpha_1\over 4\pi}\right)^2\right]
+2\vert\Lambda_2\vert^2
\left[C_2\left({\alpha_2\over 4\pi}\right)^2 
+\left({Y\over 2}\right)^2\left({\alpha_1\over 4\pi}\right)^2\right],
\label{eqe}
\end{equation}
where $C_3=4/3$ and 0 for the SU(3) triplet and singlet fields, and
$C_2=3/4$ and 0 for the SU(2) doublet and singlet fields, respectively. 
The hypercharge $Y$ is expressed as $Y=2(Q-T_3)$ by using both the electric
charge $Q$ and the diagonal SU(2) generator $T_3$. 
As it is clear from this formula for the masses of the 
scalar superpartners, we have no FCNC problem induced by these 
soft scalar masses as in the ordinary case. This is the case even if we
take account of the renormalization group effects since the
running due to the renormalization groups occurs only for the narrow range.

We apply this soft SUSY breaking scenario to the MSSM framework.
The MSSM superpotential contains the terms
\begin{equation}
W=\sum_j\left(h_j^U \hat H_2\hat Q_j \hat{\bar U_j} 
+h_j^D \hat Q_j \hat H_1\hat{\bar D_j}
+h_j^E \hat L_j \hat H_1\hat{\bar E_j} \right) 
+ \mu \hat H_1\hat H_2,
\label{eqh}
\end{equation}
where we take the Yukawa coupling diagonal basis for the quarks and 
the leptons.
All Yukawa couplings $h^f_j$ are supposed to be real.
The Higgsino mass parameter $\mu$ is generally complex.
The soft SUSY breaking terms corresponding to the superpotential
(\ref{eqh}) are introduced as\footnote{We adopt the sign convention
for $\mu, B$ and $A_f$ to make the mass eigenvalues of quarks and
leptons to be positive by a suitable field redefinition.} 
\begin{eqnarray}
-{\cal L}_{\rm soft}&=&\sum_\alpha \tilde m_\alpha^2\vert\phi_\alpha\vert^2
-\left[ \sum_j\left( A_j^U h_j^UH_2\tilde Q_j \tilde{\bar U_j}
+A_j^D h_j^D\tilde Q_j H_1\tilde{\bar D_j}
+A_j^Eh_j^E\tilde L_jH_1\tilde{\bar E_j}\right)\right. \nonumber \\ 
&-&\left. B\mu H_1H_2 -{1\over 2}\sum_r M_r\lambda_r\lambda_r 
+{\rm h.c.}\right], 
\label{eqi}
\end{eqnarray}
where we put a tilde for the superpartners of the chiral superfields
corresponding to the standard model contents.
The first term represents the soft SUSY breaking masses for
all scalar components of the MSSM chiral superfields. 
They are assumed to be given by eq.~(\ref{eqe}).
The third term in the brackets represents the
gaugino mass terms, which are supposed to be given by eq.~(\ref{eqff}). 
The soft SUSY breaking parameters $B$ and $A_j^f$ are the 
coefficients of the bilinear and trilinear scalar couplings 
with a mass dimension. 

In the minimal GMSB model, as discussed in \cite{exmgm2}, 
the soft SUSY breaking parameters $A_f$ and $B$ can be 
induced through the radiative correction.
In the case that $A_f(\Lambda)=B(\Lambda)=0$ is satisfied at the SUSY
breaking scale $\Lambda$ which is expected in many
GMSB scenario, $A_f$ and $B$ are proportional to 
$M_2$ at the low energy regions as a result of the renormalization group
effect. Thus, all of the CP phases in the soft SUSY breaking
parameters are rotated away as long as the gaugino masses are universal 
\cite{mgm2,exmgm2}.
However, in the present case this situation is broken and there remain 
the CP phases in the gaugino masses even in the case of 
$A_f(\Lambda)=B(\Lambda)=0$ since the phases in the gaugino masses are not 
universal. 
The generation of the bare $A_f$ and $B$ is completely model 
dependent in this model as in the ordinary GMSB scenario.
In the following study, we do not fix their origin and treat them 
as free parameters.

Here we make an additional assumption for the
trilinear scalar couplings such that they are proportional to the Yukawa
couplings so as to satisfy the FCNC constraints.
Although the soft SUSY breaking parameters $A_j^f$, $B$ and
$M_r$ can generally include the CP phases, all of these are not
independent physical phases. If we use the $R$-symmetry and redefine the 
fields appropriately, we can select out the physical CP phases among them.
We take them as
\begin{equation}
A_j=\vert A_j\vert e^{i\phi_{A_j}},\quad
\mu=\vert \mu\vert e^{i\phi_\mu}, \quad
M_r=\vert M_r\vert e^{i\phi_r} ~~ (r=1,3),
\label{eqii}
\end{equation}
where $B\mu$ and $M_2$ are assumed to be real.
Although the VEVs of the doublet Higgs scalars $H_1$ and $H_2$ are 
taken to be real in this definition at the tree level, the radiative 
correction could generally introduce the CP phases to them. 
Taking account of this aspect and following \cite{cphiggs1}, 
we define the VEVs of the doublet 
Higgs scalars $H_1$ and $H_2$ as
\begin{equation}
\langle H_1\rangle=\left(\begin{array}{c}v_1 \\ 0 \\ \end{array}\right),
\qquad
\langle H_2\rangle=\left(\begin{array}{c}0 \\ v_2e^{i\xi} \\ 
\end{array}\right).
\label{vev}
\end{equation}

Finally we summarize the model parameters related to the SUSY breaking.
In the present framework, the free parameters related to the masses of 
the gauginos and the scalar superpartners are confined into 
$\Lambda_1$ and $\Lambda_2$.  
Their phases are related to the physical phases $\phi_3$ and $ \phi_1$
in eq.~(\ref{eqfff}). 
Since we assume the universality for the $A$ parameter such as 
$A^f_j(\Lambda)=A$ 
at the SUSY breaking scale $\Lambda$, there remain five independent 
real parameters $\phi_\mu$, $\phi_A$, $\vert\mu\vert$, $\vert B\vert$, 
and $\vert A\vert$ in the sector of  $A^f_j$ and $B$. 
Thus, the model parameters relevant to the soft SUSY breaking are 
composed of eight real parameters,
\begin{equation}
\vert\Lambda_1\vert, \quad \vert\Lambda_2\vert, \quad
\vert A\vert, \quad \vert B\vert, \quad \vert\mu\vert,\quad
\phi_3, \quad \phi_{A}, \quad \phi_\mu. 
\label{eqrr}
\end{equation}
The phase $\xi$ in eq.~(\ref{vev}) will be determined by these parameters
through minimizing the CP-violating Higgs potential \cite{cphiggs1}.

\section{Phenomenological effects of gaugino CP phases}
\subsection{Constraints from EDMs}
In order to explain the constraints on the SUSY breaking parameters from
the EDM, we at first take a mercury case as an example to give 
a brief discussion.  The detailed discussion on the EDMs of 
the electron and the neutron 
in the present model can be found in \cite{st}. 

An effective interaction term representing the color EDM
of the quark can be written as
\begin{equation}
{\cal L}_{\rm eff}={1\over 2}~{\cal G}~\bar q {\lambda^\alpha\over 2}
\sigma_{\mu\nu}q~ F_\alpha^{\mu\nu}. 
\label{eqpp}
\end{equation}
In the estimation of the mercury EDM, we use the 
formula expressed by
\begin{equation}
d_{\rm Hg}/e=-\left(\tilde d_d-\tilde d_u-0.012\tilde d_s\right)
\times 3.2\times 10^{-2},
\end{equation}
where $\tilde d_f$ is the color EDM of an $f$-quark \cite{hg}. It is 
related to the effective coupling ${\cal G}$ in eq.~(\ref{eqpp}) 
through the formula
\begin{equation}
\tilde d_f={\rm Im}({\cal G}).
\end{equation}
The effective coupling ${\cal G}$ is composed of
the contributions from the one-loop diagrams containing one of a gluino, 
a chargino or a neutralino in the internal line.  
The experimental data for the mercury EDM $d_{\rm Hg}$ gives 
the constraint on the color EDM of the quarks such as \cite{hg1}
\begin{equation}
\left|\tilde d_d-\tilde d_u-0.012\tilde d_s\right| 
< 0.66\times 10^{-26}~{\rm cm}.
\label{merc}
\end{equation}

For the preparation to estimate the color EDM of the quarks,
we need to fix a relevant part of the MSSM to give their 
analytic formulas. 
As in the case of the EDM of the electron and the neutron,
the mixing matrices of the charginos, the neutralinos and the squarks are 
important elements to write down them at the one-loop approximation. 

In the basis of the superpotential (\ref{eqh}) and the soft 
SUSY breaking (\ref{eqi}),
the mass terms of the charginos can be written as
\begin{equation}
-\left(\tilde H_2^+, -i\lambda^+\right)
\left(\begin{array}{cc}\vert\mu\vert e^{i\phi_\mu}& \sqrt 2m_Zc_W\sin\beta\\
\sqrt 2m_Zc_W\cos\beta& M_2 \\ \end{array}\right)
\left(\begin{array}{c}\tilde H_1^- \\ -i\lambda^-\\ \end{array}\right),
\label{eqj}
\end{equation}
where $\tan\beta=v_2/v_1$ and 
the abbreviations $s_W=\sin\theta_W$ and $c_W=\cos\theta_W$ are used.
The mass eigenstates $\chi_i^\pm$ are defined in terms of the 
weak interaction eigenstates in eq.~(\ref{eqj}) 
through the unitary transformations in such a way as
\begin{equation}
\left(\begin{array}{c}\chi_1^+\\ \chi_2^+\\ \end{array}\right)
\equiv W^{(+)\dagger}\left(\begin{array}{c}\tilde H_2^+\\ -i\lambda^+\\
\end{array}\right),  \qquad
\left(\begin{array}{c} \chi_1^-\\ \chi_2^-\\ \end{array}\right)
\equiv W^{(-)\dagger}
\left(\begin{array}{c}\tilde H_1^-\\ -i\lambda^-\\ \end{array}\right).
\label{eqk}
\end{equation}
The canonically normalized neutralino basis is taken as 
${\cal N}^T=(-i\lambda_1, -i\lambda_2, 
\tilde H_1^0, \tilde H_2^0)$ and their 
mass terms are defined in such a form as
${\cal L}_{\rm mass}^{\rm n}=-{1\over 2}{\cal N}^T{\cal MN}+{\rm h.c.}$.
The 4 $\times$ 4 neutralino mass matrix ${\cal M}$ can be expressed as
\begin{equation}
\left( \begin{array}{cccc}
\vert M_1\vert e^{i\phi_1} & 0 &-m_Zs_W\cos\beta & m_Zs_W\sin\beta \\
0 & M_2 & m_Zc_W\cos\beta & -m_Zc_W\sin\beta \\
-m_Zs_W\cos\beta & m_Zc_W\cos\beta & 0 & -\vert\mu\vert e^{i\phi_\mu} \\
m_Zs_W\sin\beta & -m_Zc_W\sin\beta & -\vert\mu\vert e^{i\phi_\mu} & 0 \\
\end{array} \right).
\label{eqo}
\end{equation}
Mass eigenstates $\chi^0$ of this mass matrix are related to the weak
interaction eigenstates  ${\cal N}$ as
\begin{equation}
\chi^0\equiv U^T{\cal N},
\label{eqp}
\end{equation}
where the mass eigenvalues are defined to be real and positive so that 
the mixing matrix $U$ is considered to include the Majorana phases.

Since we do not have the flavor mixing in the sfermion sector in the
present model, the sfermion mass matrices can be reduced into 
the $2\times 2$ form for each flavor.
This $2\times 2$ sfermion mass matrix can be written in terms of 
the basis $(\tilde f_{L_\alpha}, \tilde f_{R_\alpha})$ as
\small
\begin{equation}
\left(\begin{array}{cc} |m_\alpha|^2
+\tilde m_{L_\alpha}^2+ D_{L_\alpha}^2& 
m_\alpha(\vert A_\alpha\vert e^{i \phi_{A_\alpha}}
-\vert\mu\vert e^{-i\phi_\mu} R_f)\\
m_\alpha(\vert A_\alpha\vert e^{-i\phi_{A_\alpha}}-
\vert\mu\vert e^{i\phi_\mu} R_f)& 
|m_\alpha|^2+\tilde m_{R_\alpha}^2+ D_{R_\alpha}^2\\ 
\end{array}\right),
\label{eql}
\end{equation}
\normalsize
where $m_\alpha$ and $\tilde m_{L_\alpha,R_\alpha}$ are the masses 
of the ordinary fermion 
$f_\alpha$ and its superpartners $\tilde f_{L_\alpha,R_\alpha}$, 
respectively.\footnote{In this sfermion mass matrix, the sign convention
of $A_\alpha$ is changed from the one in the previous work \cite{st}.} 
$R_f$ is $\cot\beta $ for the up component of the SU(2) fundamental
representation and $\tan\beta$ for the down component. 
$D_{L_\alpha}^2$ and $D_{R_\alpha}^2$ represent 
the $D$-term contributions, which are 
expressed as
\begin{equation}
D_{L_\alpha}^2=m_Z^2\cos 2\beta(T^f_3-Q_fs_W^2), 
\quad
D_{R_\alpha}^2=m_Z^2s_W^2Q_f\cos 2\beta, 
\label{eqm}
\end{equation}
where $T^f_3$ takes 1/2 for the sfermions in the up sector and $-1/2$ for
those in the down sector. $Q_f$ is the electric charge of the field $f$.
We define the mass eigenstates $(\tilde f_1, \tilde f_2)$ by the unitary 
transformation
\begin{equation}
\left(\begin{array}{c} \tilde f_1 \\ \tilde f_2\\ \end{array}\right)
\equiv V^{f\dagger}\left(\begin{array}{c}\tilde f_L \\ \tilde f_R\\ 
\end{array}\right).
\label{eqn}
\end{equation}

In the MSSM, there are various contributions to the quark color EDM 
$\tilde d_f$, which come from the one-loop diagram with the 
superpartners of the
standard model fields in the internal lines and can be expressed as 
$\tilde d_f\equiv\tilde d^g_f+\tilde d^\chi_f$. 
The contribution $\tilde d^g_f$ including the gluinos in the internal lines 
can be written as
\begin{eqnarray}
&&\tilde d_f^g={\alpha_s\over 8\pi}{1\over \vert M_3\vert}
\sum_{a=1}^2{\rm Im}({\cal A}_g^{f_a})\left({1\over 3}G(x_a)+3F(x_a)\right), 
\nonumber \\
&&{\cal A}_g^{f_a}=V^f_{2a}V^{f\ast}_{1a}e^{i\phi_3},
\end{eqnarray} 
where $x_a=\tilde m_a^2/\vert M_3\vert^2$.
This formula shows that the gluino phase $\phi_3$ can bring the drastic
changes in the gluino contribution to the quark color EDM. It is
remarkable that a suitable value of $\phi_3$ can change even the sign of 
the gluino contribution compared with the $\phi_3=0$ case. 

On the chargino and neutralino contributions $\tilde d^\chi_f$ 
to the quark color EDM, we can
calculate it in the same way as the ordinary EDM of the electron \cite{st}.
We find that it can be written as
\begin{eqnarray}
\hspace*{-5mm}&&\tilde d_u^\chi={\alpha m_u\over 8\pi s^2_W}
\left[{1\over m_u}\sum_{j,a}
{-2\over 3m_j}G(x_{aj}){\rm Im}({\cal A}^{u_a}_{\chi^0_j}) 
+{1\over m_W\sin\beta}\sum_{j,a}{1\over 3m_j}G(x_{aj})
{\rm Im}({\cal A}_{\chi^\pm_j}^{u_a})\right], \nonumber \\
\hspace*{-5mm}&&\tilde d_d^\chi={\alpha m_d\over 8\pi s^2_W}
\left[{1\over m_d}\sum_{j,a}
{1\over 3m_j}G(x_{aj}){\rm Im}({\cal A}^{d_a}_{\chi^0_j})
-{1\over m_W\cos\beta}\sum_{j,a}{2\over 3m_j}G(x_{aj})
{\rm Im}({\cal A}_{\chi^\pm_j}^{d_a})\right].
\end{eqnarray} 
In these formulas the mixing factors ${\cal A}^f_{\chi^\pm}$ and 
${\cal A}^f_{\chi^0}$ are defined by
\begin{eqnarray}
&&{\cal A}^{u_a}_{\chi^\pm_j}={1\over\sqrt{2}}W^{(+)}_{1j}W^{(-)}_{2j}
\vert V^d_{1a}\vert^2 +{1\over 2\cos\beta}{m_d\over m_W}
W^{(+)}_{1j}W^{(-)}_{1j}V^{d\ast}_{2a}V^d_{1a}, \nonumber \\
&&{\cal A}^{d_a}_{\chi^\pm_j}={1\over\sqrt{2}}W^{(-)}_{1j}W^{(+)}_{2j}
\vert V^u_{1a}\vert^2 +{1\over 2\sin\beta}{m_u\over m_W}
W^{(+)}_{1j}W^{(-)}_{1j}V^{u\ast}_{2a}V^u_{1a}, \nonumber \\
&&{\cal A}_{\chi^0_j}^{u_a}=-\left[\left({2\over 9}t^2_W U_{1j}^2
+{2\over 3}t_WU_{1j}U_{2j}\right)V_{1a}^{u\ast}V_{2a}^u\right. \nonumber \\
&&\hspace*{1cm}\left. -{m_u\over 2m_W\sin\beta}\left\{
\left({1\over 3}t_W U_{1j}U_{4j}+U_{2j}U_{4j}\right)\vert V_{1a}^u\vert^2
-{2\over 3}t_W U_{1j}U_{4j}\vert V_{2a}^u\vert^2\right\}\right]
\nonumber \\
&&{\cal A}_{\chi^0_j}^{d_a}=-\left[-\left({1\over 9}
t_W^2 U_{1j}^2 +{1\over 3}t_W U_{1j}U_{2j}\right)
V_{1a}^{d\ast}V_{2a}^d\right. \nonumber \\
&&\hspace*{1cm}\left.-{m_d\over 2m_W\cos\beta}\left\{\left(
{1\over 3}t_W U_{1j}U_{3j}-U_{2j}U_{3j}\right)\vert V_{1a}^d\vert^2
+{1\over 3}t_W U_{1j}U_{3j}\vert V_{2a}^d\vert^2\right\}\right],
\end{eqnarray}
where $t_W=\sin\theta_W/\cos\theta_W$.
We neglect the higher order terms of the quark mass
in the expression of ${\cal A}^f_{\chi^0}$.
Since the fermions in the external lines are very light compared with the
fields in the internal lines, $F(x)$ and $G(x)$ are 
approximately written as
\begin{equation} 
F(x)={1-3x\over (1-x)^2} -{2x^2\over (1-x)^3}\ln x,\qquad
G(x)={1+x\over (1-x)^2} +{2x\over (1-x)^3}\ln x.
\label{eqs}
\end{equation}

The gluino contribution is expected to be larger than other
contributions because of the strong coupling constant.
If we expect the cancellation among these contributions, $\tilde d^g_f$
should be suppressed to have the similar magnitude to others.
In order to find the condition for it, we may estimate a factor 
${\rm Im}({\cal A}^{f_a}_g)$ in the case of $|A|\gg|\mu|$, for example.
In that case it can be found to be approximated as
\begin{equation}
{\rm Im}({\cal A}_g^{f_a})=O\left({m_{f_a}|A|\over M_2^2}
\sin(\phi_3-\phi_A)\right). 
\label{color}
\end{equation}
This shows that the existence of the gluino phase $\phi_3$
may make it possible to suppress the gluino contribution to 
the level of others. If it happens, the experimental bounds can be satisfied. 

Both contributions of the charginos and the neutralinos are crucially
affected by the relative magnitude of $\mu$ and $M_{1,2}$.
If $|\mu|<|M_{1,2}|$ is satisfied, Higgsino components
dominate both the lightest neutralino and the lightest chargino.
Although they are expected to yield the largest contribution to the EDM,
Higgsino exchange effects can be suppressed due to the smallness of 
Yukawa couplings.
On the other hand, in the case of $|M_1|<|\mu|< |M_2|$, the lightest neutralino
and the lightest chargino seem to be dominated by the bino and the
Higgsinos, respectively. Since the gauge coupling $g_1$ is larger than
the relevant Yukawa couplings which determine the magnitude of their
contribution, the chargino contribution can be suppressed in comparison
with the neutralino contribution. As a result, they can yield the
similar order contributions. 
If the latter situation for $\mu$    
and $M_{1,2}$ is realized, the EDM constraint may be satisfied even 
in the case that the large CP phases exist in the soft SUSY breaking 
parameters. 
In the next part, we mainly focus our attention on such situations 
and carry out the numerical calculation.

\subsection{Numerical results of the EDM constraints}
At first we explain the procedure for the calculation.
We evolve the soft SUSY breaking parameters from a certain 
SUSY breaking scale $\Lambda$ to the weak scale by using 
the one-loop renormalization group equations (RGEs).
There is an ambiguity on the scale where the soft SUSY breaking
parameters are introduced and start their running. 
In the present analysis,  
we adopt $\Lambda={\rm min}~(\vert\Lambda_1\vert, \vert\Lambda_2\vert)$
as such a scale, for simplicity.
Since we mainly study the region where
$\vert\Lambda_2\vert/\vert\Lambda_1\vert$ is not so large, this 
prescription is not considered to affect the results largely.
For the gauge and Yukawa coupling constants we use the two-loop RGEs.
The RGEs from the unification scale $M_{\rm U}$ to $\Lambda$ are composed
of the SUSY ones for both the gauge and Yukawa coupling
constants. The $\beta$-functions are calculated for the MSSM contents
and the messenger fields. We solve these RGEs for various initial values 
of the Yukawa couplings at $M_U$ and examine whether the masses of the
top and bottom quarks and also the tau lepton are obtained at the weak 
scale. The messenger fields are supposed to decouple and the 
soft SUSY breaking parameters are introduced at $\Lambda$. 
Thus, the RGEs become the same as those
of the MSSM below this scale.  

In order to determine the phenomenologically interesting parameter regions,
we impose several conditions on the parameters at the weak scale 
obtained by the RGEs.
As such conditions, we adopt the followings additionally to 
the above mentioned ones:\\
(1) Various experimental mass bounds for the superpartners, such as 
gluinos, charginos, stops, staus, and charged Higgs scalars, 
should be satisfied. The color and the electromagnetic charge also should
not be broken;\\
(2) The physical true vacuum should be radiatively realized as the 
minimum of the scalar potential and satisfy
$\sin 2\beta=2B\mu/(m_1^2+m_2^2+2|\mu|^2)$. 
As another true vacuum condition, moreover, we impose 
the consistency between this $\sin 2\beta$
and the value of $\tan\beta$ predicted from the Yukawa coupling and 
the top quark mass.\footnote{We use $m_t=174.3$~GeV in this analysis.} 
Only if the difference between them is
sufficiently small, the parameters are accepted.\\
After restricting the parameter space at the high energy scale by imposing 
these conditions on the weak scale values, we finally calculate 
the EDMs of the electron, the neutron and the mercury atom. 
We compare these results with the present
experimental bounds \cite{edme,edmn}
\begin{equation}
|d_e/e|<1.6\times 10^{-27}~{\rm cm}, \qquad
|d_n/e|<0.3\times 10^{-25}~{\rm cm}
\end{equation}
for the electron and the neutron and also (\ref{merc}) for the mercury.

We present the results of the numerical 
analysis, in which we fix some
parameters to the typical values such as $|\Lambda_1|=50$~TeV,
$|\mu|=100$~GeV, and $\phi_\mu=-1.65$, for simplicity.
It seems hard to have consistent solutions 
for $|\Lambda_1|\le 35$~TeV and $|\Lambda_1|\ge 55$~TeV. 
We adopt the value of $\phi_\mu$ to introduce a seed for 
the large CP violation in the model. Since the one-loop RGEs 
do not make the phase run largely, this input value is equal to 
the weak scale one.
We also tune the initial value of $|B|$ so as to realize $\tan\beta=3.85$,
since only the very restricted values of $\tan\beta$ like $3.5 - 4$ 
seem to be consistent with the EDM constraints.  
Under these settings, we search the parameter regions which satisfy 
the above mentioned phenomenological conditions by scanning the 
remaining parameters through the following ranges at the scale $\Lambda$:
\begin{equation}
\begin{array}{cc}
50~{\rm TeV} \le |\Lambda_2| \le 150~{\rm TeV}, &
80~{\rm GeV} \le\vert A\vert\le 500~{\rm GeV},  \\
0 \le \phi_3\le \pi, &
-\pi \le \phi_A\le 0. \\ 
\end{array}
\label{region}
\end{equation}
Solutions are found for rather small values of $|A|$ such as $190 - 250$ GeV,
which satisfy $|A|>|\mu|$. The desired relation $|M_1|<|\mu|<|M_2|$
is also satisfied.

In Fig.~1 we show the allowed regions in the $(\phi_3,\phi_A)$ plane for
various values of $x(\equiv|\Lambda_2|/|\Lambda_1|)$, 
which satisfy all the EDM constraints of 
the electron, the neutron and the mercury atom. 
The imposed constraints restrict the regions of $x$ to 
$1.9~{^<_\sim}~x~{^<_\sim}~2.3$. 
Since the obtained values of $\phi_3$ yield  
small values for $\phi_1$ as found from eq.~(\ref{eqfff}), 
both sectors of the chargino and the neutralino 
seem to have no large influence of the phases in the
gaugino masses. The EDM constraint of the electron is considered 
to be satisfied without its help. 
As long as the charginos are heavier than the neutralinos, 
the cancellation between them can occur. In fact, this is satisfied in the
present solutions. 
On the other hand, the phase $\phi_3$ of the gluino mass
affects the EDMs of the neutron and the mercury atom through the gluino
contribution. 
It happens to cause the cancellation for the EDM of the neutron and
the mercury atom. In fact, the values of $\phi_3-\phi_A$ obtained here
can bring the suppression for the gluino contribution as found in 
eq.~(\ref{color}).   
This seems to suggest that the CP phases in the gaugino sector play the 
crucial role to satisfy the EDM constraints even in the case of the 
large $\phi_A$ and $\phi_\mu$.    

\input epsf
\begin{figure}[tb]
\begin{center}
\epsfxsize=8cm
\leavevmode
\epsfbox{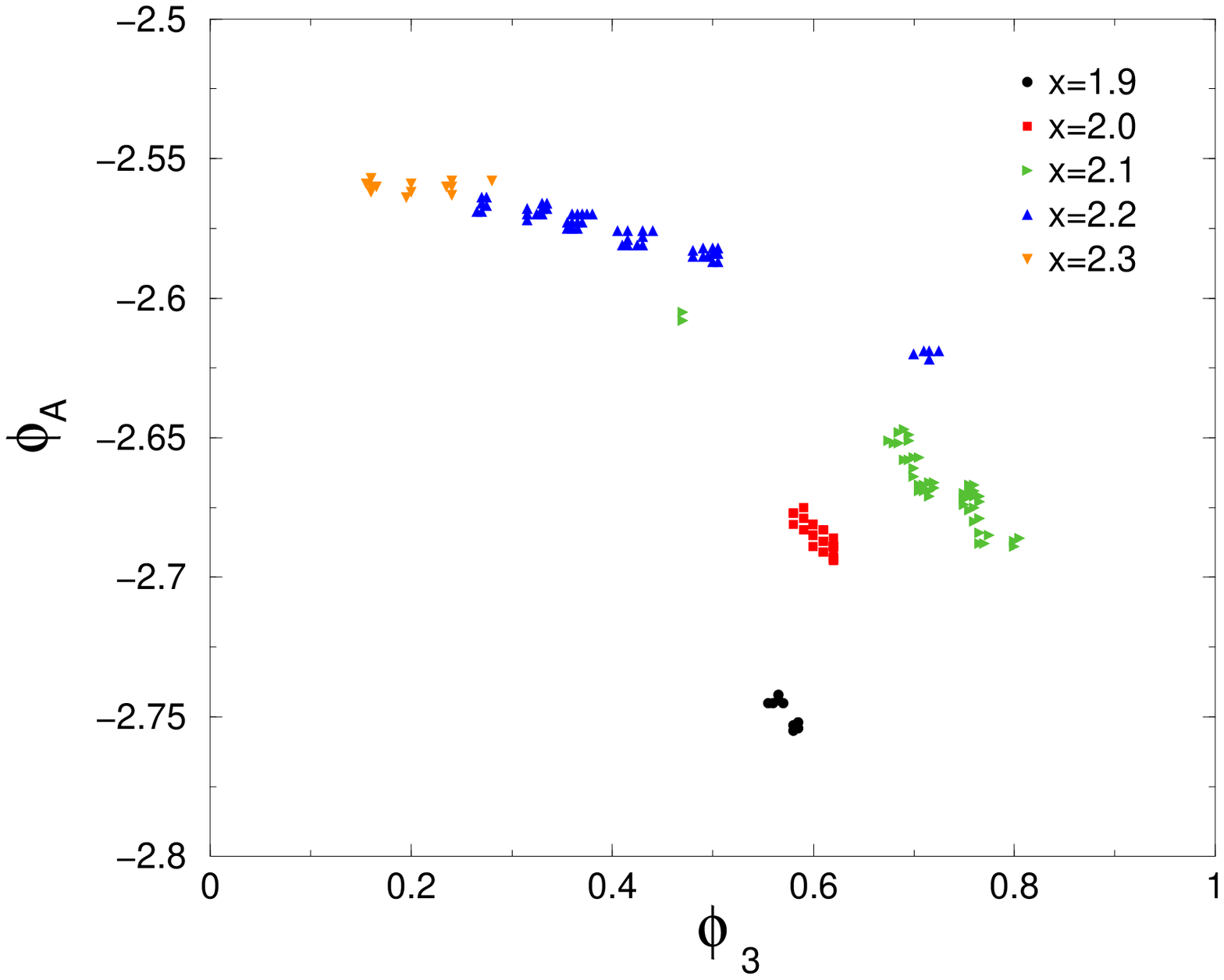}\\
\vspace*{-0.5cm}
\end{center}
{\footnotesize Fig. 1~~\  Allowed regions in the
 $(\phi_3,\phi_A)$ plane which satisfies the imposed conditions including
 the EDM constraints.}
\end{figure}

In order to show the features of the SUSY breaking for these solutions, 
we show the mass spectrum of some superpartners 
as a function of $x$ in Fig.~2.
They are determined through the values of $|\Lambda_1|$ and
$|\Lambda_2|$ as found in eqs.~(\ref{eqff}) and (\ref{eqe}). 
For the sfermion masses $\tilde m_t$, $\tilde m_b$
and $\tilde m_\tau$, we plot smaller mass eigenvalues.
The mass ratio of the chargino to the neutralino can be much larger than
that in the ordinary GMSB case $(x=1)$.
It is also remarkable that the gluino can be lighter than the squarks.
The neutralino is the lightest superpartners except for the
gravitino. The mass of the charged Higgs scalar takes 
its value in the range of $120 - 150$ GeV.

\begin{figure}[tb]
\begin{center}
\epsfxsize=8cm
\leavevmode
\epsfbox{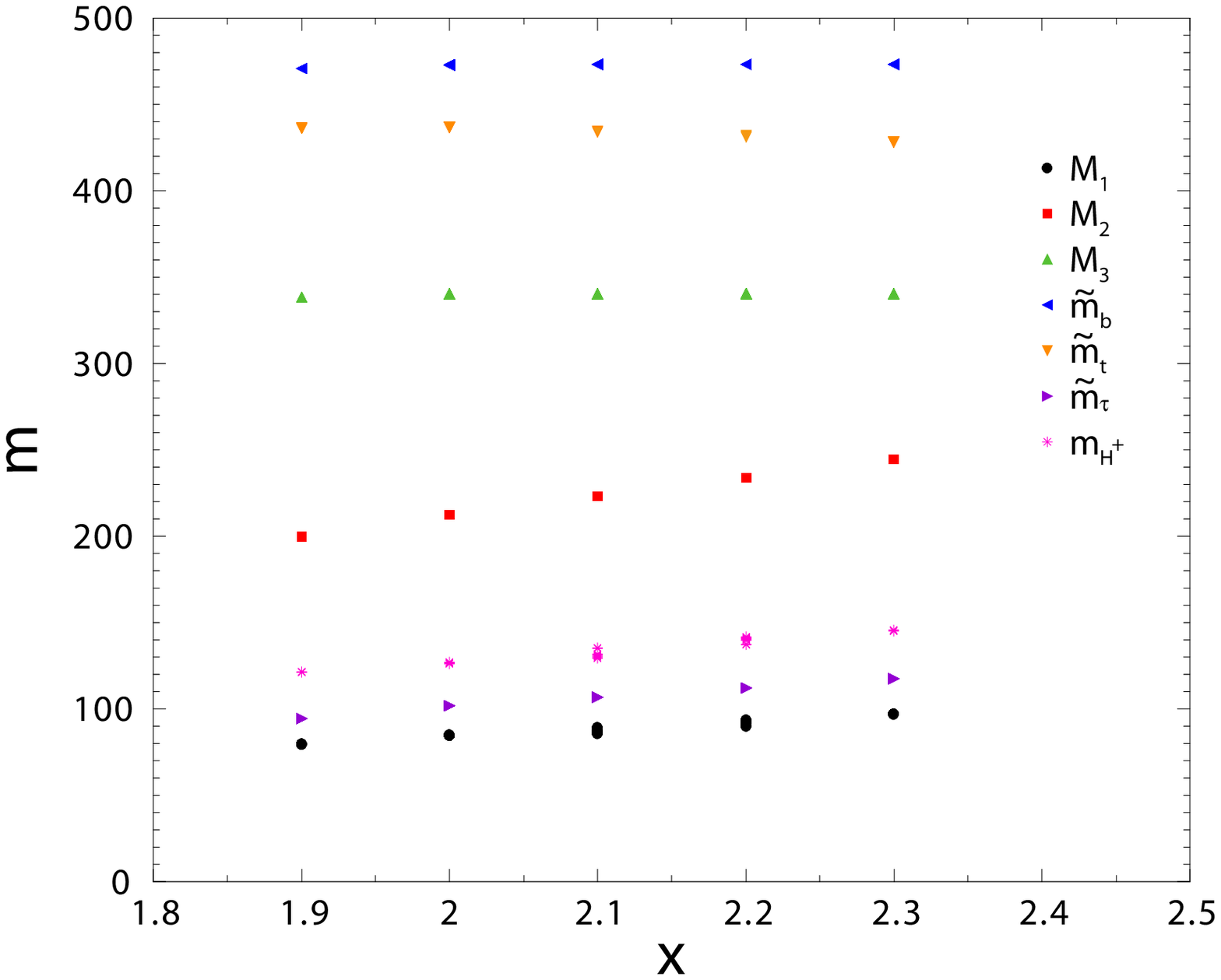}\\
\vspace*{-0.5cm}
\end{center}
{\footnotesize Fig. 2~~\  The mass spectrum of superpartners
at the weak scale as the functions of $x$. }
\end{figure}

\subsection{Phenomenology in the Higgs sector}
The allowed parameter regions obtained from the EDM constraints 
generally require small values for 
$\tan\beta$. However, as is well known in the CP conserving case, 
the small $\tan\beta$ predicts the small value of the lightest 
neutral Higgs mass in the MSSM. Then it can be a serious obstacle 
to the present solutions for the EDM constraints. 
It is an important issue to check whether our model can be
consistent with the 
constraints from the present Higgs search \cite{hgsearch}. 
In various works \cite{cphiggs0,cphiggs1,cphiggs}, it has been shown 
that the CP phases in the SUSY breaking parameters could largely
change both the Higgs mass eigenvalues and their couplings to the gauge 
bosons and the fermions. It happens due to the mixings among the CP-even and 
CP-odd Higgs scalars.
In the recent analysis of the CP violating benchmark model CPX 
with a certain top quark mass, the combined LEP 
data seem to give no universal lower bound 
for the lightest neutral Higgs mass, although they can restrict 
the $\tan\beta$ to be larger than 2.6 \cite{hgsearch}.
In the present model, the similar feature may also be found for 
the parameter region derived from the EDM constraints, and
it can be consistent with the present experimental data for the Higgs sector.

In order to study this aspect, we follow the one-loop effective
potential method discussed in \cite{cphiggs1}, in which the one-loop 
effective potential is expanded by the operators up to the fourth 
order and the effective Higgs quartic couplings are analytically determined.
Our EDM study suggests that the small $\tan\beta$ is favorable 
and also both $|A|$ and $|\mu|$ tend to be smaller than the soft 
scalar masses of the left- and right-handed stops. 
These features seems to make the usage of this 
method validate for the present analysis. 
In our model the gaugino masses are non-universal 
and then there can be physical CP phases in the gaugino masses in
addition to those in the $\mu$ and $A$-parameters.
This is the different situation from that in \cite{cphiggs1}. 
The gaugino phases could contribute to the one-loop effective
potential mainly through the neutralino and chargino loops.
However, since these CP violating corrections to the effective potential are 
considered to be smaller than the one coming from the stop contribution, 
we neglect them in this study,  and we directly apply the formulas 
in \cite{cphiggs1} to this analysis.

In the following part, we focus our study on the mass eigenvalues of 
the Higgs scalars and
the couplings between the Higgs scalars and the gauge bosons. 
They can be represented by using the Higgs quartic 
effective couplings $\lambda_{1-7}$. These definitions and
their analytical formulas \cite{cphiggs1} are presented in the appendix.
If we impose the potential minimum conditions, we can write the neutral
Higgs mass matrix in the form as 
\begin{equation}
{\cal M}_0^2=\left(\begin{array}{ccc}
{\cal M}_S^2 & ({\cal M}_{PS}^2)^T \\
{\cal M}_{PS}^2 & M_a^2 \\ 
\end{array}\right),
\end{equation}
where ${\cal M}_S^2$ is a $2\times 2$ mass matrix for the CP-even Higgs
scalars and ${\cal M}_{PS}^2$ is a $1\times 2$ matrix representing the
mixing among the CP-odd and CP-even Higgs scalars. 
These sub-matrices can be expressed as
\begin{eqnarray}
&&\hspace*{-7mm}{\cal M}_S^2=M_a^2\left(\begin{array}{cc}
s_\beta^2 & -s_\beta c_\beta \\ -s_\beta c_\beta & c_\beta^2 \\
\end{array}\right) +~ 2v^2\times  \nonumber\\
&&\hspace*{-4mm}
\small
\left(\begin{array}{cc}
-2(\lambda_1c^2_\beta+{\rm Re}(\lambda_5e^{2i\xi})s_\beta^2
+{\rm Re}(\lambda_6e^{i\xi})s_\beta c_\beta)&
\lambda_{34}s_\beta c_\beta+
{\rm Re}(\lambda_6e^{i\xi})c_\beta^2+{\rm Re}(\lambda_7e^{i\xi})s_\beta^2\\
\lambda_{34}s_\beta c_\beta+
{\rm Re}(\lambda_6e^{i\xi})c_\beta^2+{\rm Re}(\lambda_7e^{i\xi})s_\beta^2 &
-2(\lambda_2s^2_\beta+{\rm Re}(\lambda_5e^{2i\xi})c_\beta^2
+{\rm Re}(\lambda_7e^{i\xi})s_\beta c_\beta)\\ 
\end{array}\right), \nonumber\\
&&\hspace*{-7mm}
\normalsize
{\cal M}_{PS}^2=2v^2\left(\begin{array}{cc} 
{\rm Im}(\lambda_5e^{2i\xi})s_\beta+{\rm Im}(\lambda_6e^{i\xi})c_\beta &
{\rm Im}(\lambda_5e^{2i\xi})c_\beta+{\rm Im}(\lambda_7e^{i\xi})s_\beta \\
\end{array} \right), 
\end{eqnarray}
where $\lambda_{34}=\lambda_3+\lambda_4$, $s_\beta=\sin\beta$ and
  $c_\beta=\cos\beta$.
$M_a^2$ corresponds to 
the physical mass of the CP-odd Higgs 
scalar in the CP conserving MSSM, and it can be written as
\begin{equation}
M_a^2={1\over s_\beta c_\beta}\left[{\rm Re}(m_{12}^2 e^{i\xi})+2v^2\left\{
2{\rm Re}(\lambda_5 e^{2i\xi})s_\beta c_\beta+{1\over 2}
{\rm Re}(\lambda_6 e^{i\xi})c_\beta^2+{1\over 2}
{\rm Re}(\lambda_7e^{i\xi})s_\beta^2\right\}\right].
\end{equation}
The mass of the charged Higgs scalars can be expressed as
\begin{eqnarray}
M_{H^\pm}^2&=&{1\over s_\beta c_\beta}\Big[{\rm Re}(m_{12}^2 e^{i\xi})
\nonumber\\
&+&\left.2v^2\left\{{1\over 2}\lambda_4s_\beta c_\beta+
{\rm Re}(\lambda_5 e^{2i\xi})s_\beta c_\beta+
{1\over 2}{\rm Re}(\lambda_6 e^{i\xi})c_\beta^2+
{1\over 2}{\rm Re}(\lambda_7e^{i\xi})s_\beta^2\right\}\right].
\end{eqnarray}

The Higgs couplings to the gauge bosons are also changed from 
those in the CP conserving case.
This occurs due to the mixing among the
CP-even and CP-odd Higgs scalars, which is induced by ${\cal M}_{PS}^2$. 
The interaction Lagrangian for the mass eigenstates of the Higgs scalars
$H_i$ is found to be expressed as  
\begin{eqnarray}
&&{\cal L}_{HVV}=g_2M_W\sum_{i=1}^3g_{{H_i}VV}\left(H_i
W_\mu^+W^{-\mu}+{1\over 2c_W^2}H_i Z_\mu Z^\mu\right), \nonumber \\
&&{\cal L}_{HHZ}={g_2\over 2c_W}\sum_{j>i=1}^3g_{H_iH_jZ}\left(H_i
\stackrel{\leftrightarrow}{\partial}_\mu H_j\right)Z^\mu, \nonumber\\
&&{\cal L}_{HH^\pm W^\mp}={g_2\over 2}\sum_{i=1}^3\left[g_{H_iH^-W^+}
\left(H_i i\stackrel{\leftrightarrow}{\partial}_\mu H^-\right)W^{+\mu}
+ h.c.\right].
\label{coupling}
\end{eqnarray}
In these interaction Lagrangians for the Higgs scalars,  
each coupling normalized to the value in the standard model can be written as
\begin{eqnarray}
&&g_{H_iVV}=c_\beta O_{1i}+s_\beta O_{2i}, \qquad (V=W^\pm,Z)\nonumber \\
&&g_{H_iH_jZ}=O_{3i}(c_\beta O_{2j}-s_\beta O_{1j})
-O_{3j}(c_\beta O_{2i}-s_\beta O_{1i}), \nonumber\\
&&g_{H_iH^-W^+}= c_\beta O_{2i}-s_\beta O_{1i}+iO_{3i}, 
\end{eqnarray}
where $O_{ij}$ is the element of the orthogonal matrix which relates the
mass eigenstates $H_i$ to the weak eigenstates. 
It is defined as the diagonalization matrix for ${\cal M}_0^2$ in such a way as
\begin{equation}
O^T{\cal M}_0^2 O={\rm diag}(m_{H_1}^2,m_{H_2}^2,m_{H_3}^2),
\end{equation}
where the mass eigenvalues $m_{H_i}^2$ for the eigenstates $H_i$ satisfy 
the relation such as $m_{H_1}^2\le m_{H_2}^2\le m_{H_3}^2$.
Since there are the following relations among these neutral Higgs couplings:
\begin{equation}
g_{H_kVV}=\varepsilon_{ijk}g_{H_iH_jZ}, \qquad \sum_{i=1}^3g_{H_iVV}^2=1,
\label{unit}
\end{equation}
all of the couplings of the neutral Higgs scalars to the gauge bosons can be 
completely determined by the two values of $g_{H_iZZ}$, 
for example \cite{ghcoup}.

\begin{figure}[tb]
\begin{center}
\epsfxsize=8cm
\leavevmode
\epsfbox{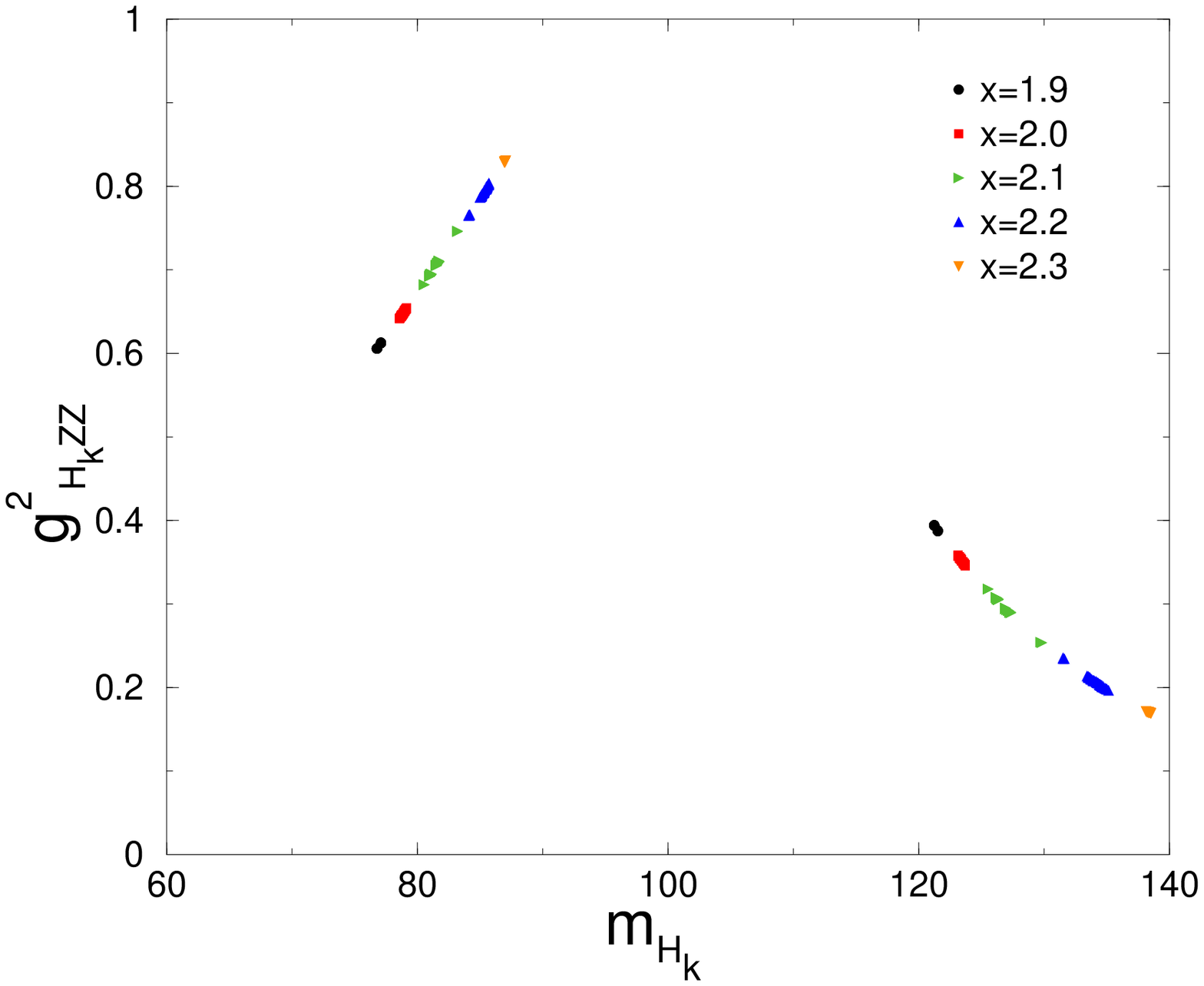}
\end{center}
{\footnotesize Fig. 3~~\  The mass eigenvalues and the coupling
 constants with the gauge bosons of the neutral Higgs scalars $H_1$ and $H_3$.}
\end{figure}

As mentioned already, the CP violating effect in the Higgs sector 
appears through the mixing ${\cal M}_{PS}^2$ between the CP-even 
and CP-odd Higgs scalars.
If we use the analytic formulas for the quartic couplings 
$\lambda_{5,6,7}$ in the appendix,
we find that the order of these off-diagonal elements are estimated as
\begin{equation}
{\cal M}_{SP}^2\simeq O\left({m_t^4\over v^2}
{{\rm Im}(A\mu)\over 64\pi^2M_S^2}\right)
=O\left({v^2|A||\mu|\sin\phi_{CP}\over 64\pi^2M_S^2}
\left(\tan^2\beta\over 1+\tan^2\beta \right)^2\right),
\end{equation}
where $\phi_{CP}=\phi_A+\phi_\mu$ which is the measure for the CP
violation in the Higgs sector. 
If ${\rm Im}(A\mu)$ can have large values,
they can be so large as to have crucial
effects on the composition of the mass eigenstates of the neutral Higgs
scalars. Thus, the larger values of $\vert\mu\vert$, $\vert A\vert$ and $\phi_{CP}$
constitute the interesting parameter regions, in which the CP 
violating effects on the Higgs sector are substantial. 
On the other hand, as discussed in \cite{cphiggs1}, 
the CP violating effects on the Higgs 
sector also tend to be enhanced in the case that
the charged Higgs mass $M_{H^\pm}$ takes a small value. \footnote{This 
is expected to be realized for the case of small value of $m_{12}^2(=B\mu)$.
However, if the top quark mass is larger, the CP violating effect
seems to appear independently of the charged Higgs mass \cite{hgsearch}.} 
In the present model, $A$, $B$ and $\mu$ are free parameters. 
Since they are not directly related to other SUSY breaking 
parameters such as the gaugino masses and the sfermion masses,
we can study their interesting regions without 
making large influence on the mass spectrum of the gauginos and 
the sfermions as long as $\tilde m_f>|A|, |\mu|$ is satisfied.
However, we should note that the EDM constraints tend to favor the small
values of $\tan\beta$, $|A|$ and $|\mu|$ as partially seen 
in eq.~(\ref{color}), for example. 
Thus, the EDM constraints may make the CP violating effects 
in the Higgs sector small even in the
case with the large $\phi_{CP}$. Although the cases where $|A|$ and
$|\mu|$ are not large but $\phi_{\rm CP}$ is $O(1)$ seem to be promising
in the present context, the situation is subtle and the
detailed numerical study is required to clarify this point.

\begin{figure}[tb]
\begin{center}
\epsfxsize=8cm
\leavevmode
\epsfbox{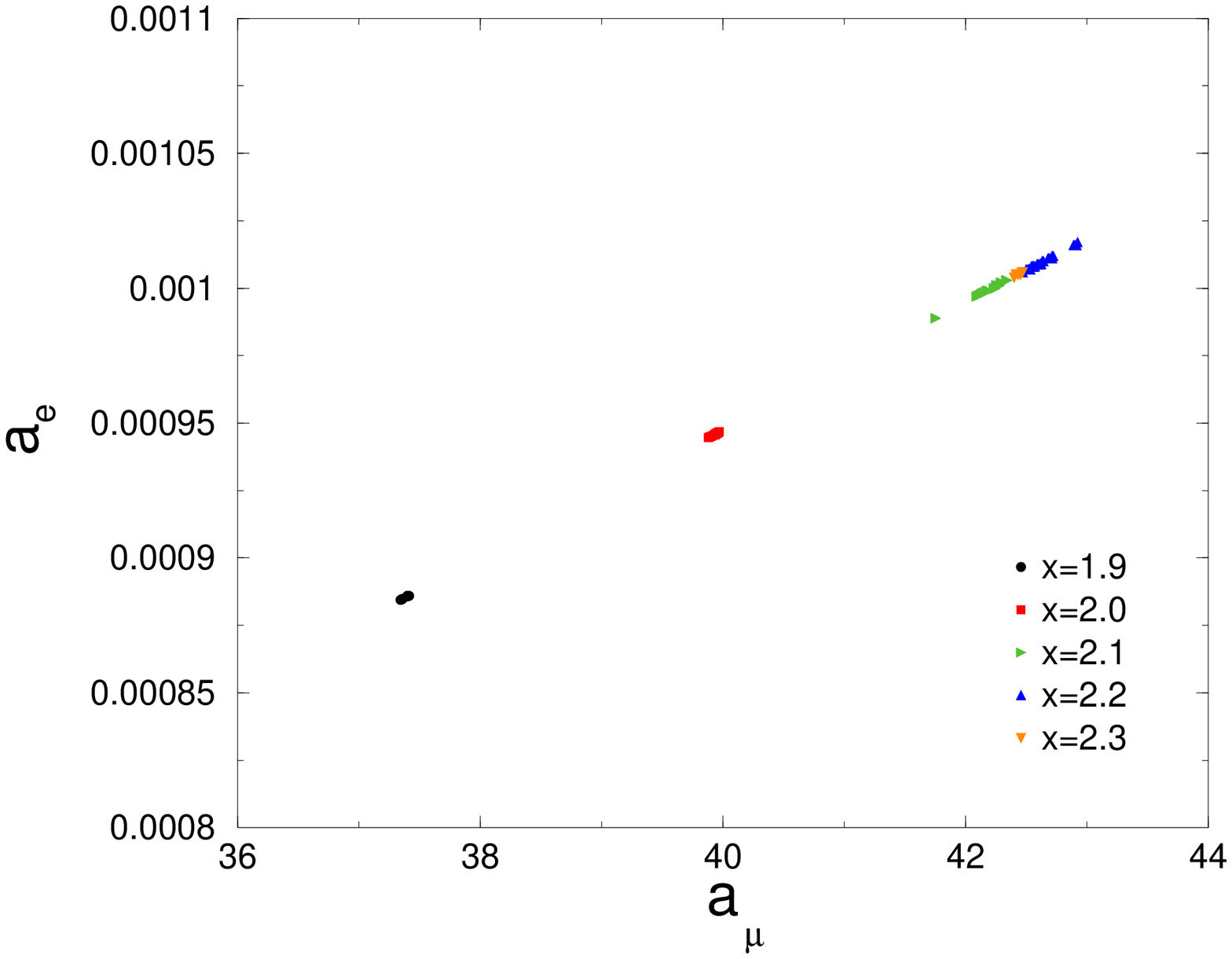}\\
\end{center}
\vspace*{-0.5cm}
{\footnotesize Fig. 4~~\  The SUSY contributions to the $g-2$ of
 the muon and the electron expected for the obtained solutions.}
\end{figure}

We calculate both the Higgs mass eigenvalues and the Higgs couplings for the
parameter sets obtained in the previous part.
In Fig.~3 we plot the mass eigenvalues of the neutral 
Higgs scalars $H_k$ and their
coupling constants $g_{H_kZZ}^2$ with the gauge bosons.
Both the lightest neutral Higgs scalar $H_1$ and the heaviest 
one $H_3$ are plotted in the same figure for each value of $x$.
Since eq.~(\ref{unit}) is satisfied among the couplings,
$g^2_{H_2ZZ}$ is negligible in the present case. 
The mass eigenvalues $m_{H_k}$ are the increasing functions of $x$.
If we combine this figure with Fig.~1, 
we can see that they are affected largely by the phase $\phi_{CP}$.
Since these Higgs mass eigenvalues take rather small values, 
the model might be considered to have been already excluded by the Higgs 
search at the LEP.
However, the Higgs couplings are also influenced largely by this $\phi_{CP}$. 
as observed in Fig.~3. 
Figure 3 shows that the $H_1$ coupling $g^2_{H_1ZZ}$ can be 
much smaller than the MSSM one.
The values of $m_{H_1}$ and $g^2_{H_1ZZ}$ shown in Fig.~3 seem to be
marginal against the LEP2 data \cite{hgsearch}. 

We could only say that our solutions for the EDM 
constraints might be consistent with the present Higgs phenomenology 
on the basis of our analysis.
However, our results suggest that the validity of the model can be
checked if the new experiments start at the LHC, anyway. 
Our analysis can also give several predictions for the relevant 
physical quantities. As a good example, we estimate the SUSY contributions 
$a_\mu$ and $a_e$ to the anomalous magnetic moment 
of the muon and the electron.
The results are shown in Fig.~4. Both of $a_\mu$ and $a_e$ are
plotted in the unit of $10^{-11}$. This predicted values of the muon $g-2$ 
seem to be in the interesting regions for the present experimental data.
 
\section{Summary}
Non-universality of the gaugino masses can potentially cause 
various interesting phenomenology at the weak scale.
We have considered the extended gauge mediation SUSY
breaking scenario as an concrete example which could realize the
non-universal gaugino masses. 
In this model the CP phases can remain in the gaugino sector as the 
physical phases after the $R$-transformation. 
In addition to this aspect, the model has several features different 
from the usual MSSM or the ordinary gauge mediation SUSY breaking. 
For example, the SU(2)$_L$ non-singlet superpartners tend to be heavier 
than the SU(2)$_L$ singlet ones whether they are colored or not. 
The right-handed stop becomes rather light and the neutralino can 
be lighter than the stau. 
These features can affect various phenomenology to give the 
different results from the ordinary MSSM.

We have calculated the effect of the CP phases on the EDM of the mercury 
atom, the electron and the neutron by solving the RGEs for the soft 
SUSY breaking parameters. As a result of this analysis,
we have found that the experimental bounds for these EDMs could be 
simultaneously satisfied without assuming the heavy superpartners 
with the mass of $O(1)$~TeV even in the case that the soft SUSY breaking 
parameters have the large CP phases.
The effective cancellation among the contributions from the gluino, the
neutralino, and the chargino makes them possible to satisfy the
experimental constraints. In this cancellation, the CP
phases in the gaugino sector seem to play the crucial role. 
Although this kind of phenomena have already been suggested in several 
works, we have shown this in the concrete model with the definite
spectrum of the superpartners.

The Higgs sector could also be affected by the the existence of
the large CP phases in the soft SUSY breaking parameters.
Since the CP-even Higgs scalars mix with the CP-odd Higgs 
scalar, the lightest neutral Higgs mass and its couplings to the gauge
bosons could be largely modified from those in the CP invariant case. 
We have studied these aspects in the parameter regions where the EDM
constraints are satisfied. From this study, we have found that our model
might be consistent with the present data obtained from the Higgs search
at the LEP2. The validity of the model will be checked at the LHC.

\vspace*{.5cm}
\noindent
The author thanks Dr. H.~Tsuchida for the collaboration at the first
stage of the numerical study.
This work is supported in part by a Grant-in-Aid for Scientific 
Research (C) from Japan Society for Promotion of Science
(No.~14540251, 17540246).

\newpage
\noindent
{\Large\bf Appendix}

\noindent
The effective Lagrangian which describes the most general CP-violating
Higgs potential of the MSSM is given by 
\begin{eqnarray}
{\cal L}&=&\mu^2_1(\Phi_1^\dagger\Phi_1)+\mu_2^2(\Phi_2^\dagger\Phi_2)
+m_{12}^2(\Phi_1^\dagger\Phi_2)+m_{12}^{\ast 2}(\Phi_2^\dagger\Phi_1)
+\lambda_1(\Phi_1^\dagger\Phi_1)^2+\lambda_2(\Phi_2^\dagger\Phi_2)^2
\nonumber \\
&+&\lambda_3(\Phi_1^\dagger\Phi_1)(\Phi_2^\dagger\Phi_2)
+\lambda_4(\Phi_1^\dagger\Phi_2)(\Phi_2^\dagger\Phi_1)
+\lambda_5(\Phi_1^\dagger\Phi_2)^2 +\lambda_5^\ast(\Phi_2^\dagger\Phi_1)^2
\nonumber\\
&+&\lambda_6(\Phi_1^\dagger\Phi_1)(\Phi_1^\dagger\Phi_2)
+\lambda_6^\ast(\Phi_1^\dagger\Phi_1)(\Phi_2^\dagger\Phi_1)
+\lambda_7(\Phi_2^\dagger\Phi_2)(\Phi_1^\dagger\Phi_2) 
+\lambda_7^\ast(\Phi_2^\dagger\Phi_2)(\Phi_2^\dagger\Phi_1),
\label{lag}
\end{eqnarray}
where $\Phi_{1,2}$ are related to the scalar components $H_{1,2}$ of
the Higgs superfields $\hat H_{1,2}$ through 
$H_1=i\tau_2\Phi_1^\ast$ and $H_2=\Phi_2$. At the tree level,
coefficients in eq.~(\ref{lag}) are represented as
\begin{eqnarray}
&&\mu_1^2=-m_1^2-\vert\mu\vert^2, \quad \mu_2^2=-m_2^2-\vert\mu\vert^2, \quad
m_{12}^2=B\mu, \nonumber\\
&&\lambda_1=\lambda_2=-{1\over 8}(g_2^2+g_1^2), \quad
\lambda_3=-{1\over 4}(g_2^2-g_1^2), \quad \lambda_4={1\over 2}g_2^2, 
\nonumber\\
&&\lambda_5=\lambda_6=\lambda_7=0.
\end{eqnarray}  
Taking account of radiative corrections due to the trilinear Yukawa
couplings between the Higgs scalars and stops/sbottoms, 
the quartic couplings $\lambda_{5,6,7}$ generally have complex 
nonzero values. 
If we assume that $M_S$ is a SUSY breaking scale,
analytic expressions of these quartic couplings are
given by \cite{cphiggs1},
\begin{eqnarray}
\lambda_1&=&-{g_2^2+g_1^2\over 8}\left(1-{3\over 8\pi^2}h_b^2t\right) 
\nonumber\\
&&-{3\over 16\pi^2}h_b^4\left[t+{1\over 2}X_b
+{1\over 16\pi^2}\left({3\over2}h_b^2+{1\over 2}h_t^2
-8g_3^2\right)\left(X_bt+t^2\right)\right] \nonumber \\
&&+{3\over 192\pi^2}h_t^2{\vert\mu\vert^4\over M_S^4}
\left[1+{1\over 16\pi^2}\left(9h_t^2-5h_b^2-16g_3^2\right)t\right], \nonumber\\
\lambda_2&=&-{g_2^2+g_1^2\over 8}\left(1-{3\over 8\pi^2}h_t^2t\right) 
\nonumber\\
&&-{3\over 16\pi^2}h_t^4\left[t+{1\over 2}X_t
+{1\over 16\pi^2}\left({3\over2}h_t^2+{1\over 2}h_b^2
-8g_3^2\right)\left(X_tt+t^2\right)\right] \nonumber\\
&&+{3\over 192\pi^2}h_b^2{\vert\mu\vert^4\over M_S^4}
\left[1+{1\over 16\pi^2}\left(9h_b^2-5h_t^2-16g_3^2\right)t\right], \nonumber\\
\lambda_3&=&-{g_2^2-g_1^2\over 8}\left[1-{3\over 16\pi^2}(h_t^2+h_b^2)t\right] 
\nonumber\\
&&-{3\over 8\pi^2}h_t^2h_b^2\left[t+{1\over 2}X_{tb}
+{1\over 16\pi^2}\left(h_t^2+h_b^2-8g_3^2\right)\left(X_{tb}t+t^2\right)\right]
\nonumber\\
&&-{3\over 96\pi^2}h_t^4\left({3\vert\mu\vert^2\over M_S^2}
-{\vert\mu\vert^2\vert A_t\vert^2\over M_S^4}\right)
\left[1+{1\over 16\pi^2}\left(6h_t^2-2h_b^2-16g_3^2\right)t\right], \nonumber\\
&&-{3\over 96\pi^2}h_b^4\left({3\vert\mu\vert^2\over M_S^2}
-{\vert\mu\vert^2\vert A_b\vert^2\over M_S^4}\right)
\left[1+{1\over 16\pi^2}\left(6h_b^2-2h_t^2-16g_3^2\right)t\right], \nonumber\\
\lambda_4&=&{g_2^2\over 2}\left[1-{3\over 16\pi^2}(h_t^2+h_b^2)t\right] 
\nonumber\\
&&+{3\over 8\pi^2}h_t^2h_b^2\left[t+{1\over 2}X_{tb}
+{1\over 16\pi^2}\left(h_t^2+h_b^2-8g_3^2\right)\left(X_{tb}t+t^2\right)\right]
\nonumber\\
&&-{3\over 96\pi^2}h_t^4\left({3\vert\mu\vert^2\over M_S^2}
-{\vert\mu\vert^2\vert A_t\vert^2\over M_S^4}\right)
\left[1+{1\over 16\pi^2}\left(6h_t^2-2h_b^2-16g_3^2\right)t\right], \nonumber\\
&&-{3\over 96\pi^2}h_b^4\left({3\vert\mu\vert^2\over M_S^2}
-{\vert\mu\vert^2\vert A_b\vert^2\over M_S^4}\right)
\left[1+{1\over 16\pi^2}\left(6h_b^2-2h_t^2-16g_3^2\right)t\right], \nonumber\\
\lambda_5&=&{3\over 192\pi^2}h_t^4
{\mu^2A_t^2\over M_S^4}
\left[1+{1\over 16\pi^2}\left(6h_t^2-2h_b^2-16g_3^2\right)t\right], \nonumber\\
&&+{3\over 192\pi^2}h_b^4{\mu^2A_b^2\over M_S^4}
\left[1+{1\over 16\pi^2}\left(6h_b^2-2h_t^2-16g_3^2\right)t\right], \nonumber\\
\lambda_6&=&-{3\over 96\pi^2}h_t^4
{\vert\mu\vert^2\mu A_t\over M_S^4}
\left[1+{1\over 16\pi^2}\left({15\over 2}h_t^2-{7\over 2}h_b^2
-16g_3^2\right)t\right], \nonumber\\
&&+{3\over 96\pi^2}h_b^4{\mu\over M_S}
\left({6A_b\over M_S}-{\vert A_b\vert^2A_b\over M_S^3}\right)
\left[1+{1\over 16\pi^2}\left({9\over 2}h_b^2-{1\over 2}h_t^2
-16g_3^2\right)t\right], \nonumber\\
\lambda_7&=&-{3\over 96\pi^2}h_b^4
{\vert\mu\vert^2\mu A_b\over M_S^4}
\left[1+{1\over 16\pi^2}\left({15\over 2}h_b^2-{7\over 2}h_t^2
-16g_3^2\right)t\right], \nonumber\\
&&+{3\over 96\pi^2}h_t^4{\mu\over M_S}
\left({6A_t\over M_S}-{\vert A_t\vert^2A_t\over M_S^3}\right)
\left[1+{1\over 16\pi^2}\left({9\over 2}h_t^2-{1\over 2}h_b^2
-16g_3^2\right)t\right], 
\end{eqnarray}
In these formulas, the following definitions are used:
\begin{eqnarray}
&&t=\ln\left({M_S^2\over \bar{m}_t^2}\right), \qquad
h_t={m_t(\bar{m}_t)\over v\sin\beta}, \qquad 
h_b={m_b(\bar{m}_t)\over v\cos\beta}, \nonumber \\
&&X_t={2\vert A_t\vert^2\over M_S^2}
\left(1-{\vert A_t\vert^2 \over 12M_S^2}\right), \qquad
X_b={2\vert A_b\vert^2 \over M_S^2}
\left(1-{\vert A_b\vert^2 \over 12M_S^2}\right), \nonumber \\
&&X_{tb}={\vert A_t\vert^2+\vert A_b\vert^2 +2{\rm Re}(A_b^\ast A_t)
\over 2M_S^2}-{\vert\mu\vert^2\over M_S^2}
-{\vert\vert\mu\vert^2-A_b^\ast A_t\vert^2 \over 6M_S^4},
\end{eqnarray}
where $\bar{m}_t$ is the pole mass of the top quark, which can be
related to the running mass $m_t$ as
\begin{equation}
m_t(\bar{m}_t)={\bar{m}_t\over 1+{4\over 3\pi}\alpha_3(\bar{m}_t)}.
\end{equation} 
We assume that the SUSY breaking scale $M_S^2$ is defined as the
arithmetic average of the squared stop mass eigenvalues 
in the numerical calculation of the Higgs sector.


\end{document}